\documentclass[aps,prl,twocolumn,showpacs,superscriptaddress,groupedaddress,nofootinbib]{revtex4-1} 
\usepackage{graphicx}
\usepackage{amsfonts,amssymb}
\usepackage{amsmath}

\begin{document}

\title{Complex Systems Science: Dreams of Universality, Reality of Interdisciplinarity}

\author{S\'ebastian Grauwin$^{1,2,3}$, Guillaume Beslon$^{1,2,4}$, Eric Fleury$^{1,2,5,6}$, Sara Franceschelli$^{1,2,6}$, C\'eline Robardet$^{1,2,4}$, Jean-Baptiste Rouquier$^{7}$, Pablo Jensen$^{1,2,3,\ast}$} 
\address{
\begin{center}
\begin{minipage}{12cm}
\begin{enumerate}
\item Universit\'e de Lyon, France
\item IXXI, Rh\^one Alpes Institute of complex systems, 69364 Lyon, France
\item Laboratoire de Physique, UMR CNRS 5672, ENS de Lyon, 69364 Lyon, France
\item LIRIS, UMR CNRS 5205, INSA-Lyon, 69621 Villeurbanne, France
\item LIP, UMR CNRS 5668, INRIA and ENS de Lyon, 69364 Lyon, France
\item ENS de Lyon, 69364 Lyon, France
\item ISC-PIF, Paris, France
\end{enumerate}
$\hspace{5mm}\ast$ E-mail: pablo.jensen@ens-lyon.fr
\end{minipage}
\end{center}
}

\maketitle

{\bf Using a large database ($\sim$ 215 000 records) of relevant articles, we empirically study the ``complex systems'' field and its claims to find universal principles applying to systems in general. The study of references shared by the papers allows us to obtain a global point of view on the structure of this highly interdisciplinary field. We show that its overall coherence does not arise from a universal theory but instead from computational techniques and fruitful adaptations of the idea of self-organization to specific systems. We also find that communication between different disciplines goes through specific ``trading zones'', {\it ie} sub-communities that create an interface around specific tools (a DNA microchip) or concepts (a network). }

\section*{Introduction}
Fundamental science has striven to reduce the diversity of the world to some stable building blocks such as atoms and genes. To be fruitful, this reductionist approach must be complemented by the reverse step of obtaining the properties of the whole (materials, organisms) by combining the microscopic entities, a notoriously difficult task \cite{nature_complex,anderson,pnas_schelling,gannon}. The science of complex systems tackles this challenge, albeit from a different perspective. It adds the idea that ``universal principles'' could exist, which would allow for the prediction of the organization of the whole regardless of the nature of the microscopic entities. Ludwig Von Bertalanffy wrote already in 1968: ``{\it It seems legitimate to ask for a theory, not of systems of a more or less special kind, but of universal principles applying to systems in general }'' \cite{gst}. This dream of universality is still active: ``{\it [Complex networks science] suggests that nature has some universal organizational principles that might finally allow us to formulate a general theory of complex systems }'' \cite{universal_cn}. Have such universal principles been discovered? Could they link disciplines such as sociology, biology, physics and computer science, which are very different in both methodology and objects of inquiry \cite{santafe}?

\section*{Results}

In this paper, we empirically study the ``complex systems'' field using the quantitative tools developed to understand the organization of scientific fields \cite{Small99} and their evolution \cite{glanzel,cointet,mogoutov}. Global science maps \cite{Small99,klavans,Small73,borner,borneratlas,noyons,Leydesdorff,rafols,Skupin,cobo,leydesgeo,grauwin} have become feasible recently, offering a tentative overall view of scientific fields and fostering dreams of a ``science of science'' \cite{borner}. Specifically, to collect a representative database of articles, we selected from the ISI Web of knowledge \cite{wos} all records containing topic keywords relevant for the field of complex systems (Table \ref{TK}). Table \ref{mostusedrefs} contains the 20 most frequent cited references and journals within our dataset. To analyze the data, we build a network \cite{borner} in which the $\sim$ 215 000 articles are the nodes. These nodes are linked according to the proportion of shared references (bibliographic coupling \cite{kessler}). For this study, bibliographic coupling offers two advantages over the more usual co-citation link: it offers a faithful representation of the fields, giving equal weight to all published papers (whether cited or not) and it can be applied to recent papers (which have not yet been cited). For more details, the reader is referred to the section ``Methods''. 

\begin{table}[h!]
\caption{
{\bf Topic keywords used in our request in the ISI Web of Knowledge database and number of articles matching independently to each of these topic keywords.} Each topic keywords except the first six where coupled with the topic keywords ``complex*''. We moreover rejected the articles containing the topic keywords ``complex scaling'' or ``linear search'', two terms refering respectively to (heavily used) specific methods of quantum chemistry and computer science.}
\scriptsize{
\begin{tabular}{lr}
\hline
topic keywords & Results \\
\hline
self organ* & $32484$ \\
complex network* & $6953$ \\
dynamical system & $8205$ \\
econophysics & $633$ \\
strange attractor & $769$ \\
synergetics & $379$ \\
adaptive system* & $1141$ \\
artificial intelligence & $1812$ \\
attractor & $1034$ \\
bifurcation & $3164$ \\
chaos  & $5370$ \\
control  & $116017$ \\
criticality & $980$ \\
ecology & $5869$ \\
economics & $2243$ \\
epistemology & $345$ \\
far from equilibrium & $253$ \\
feedback & $12881$ \\
fractal & $3867$ \\
ising & $975$ \\
multi agent & $2032$ \\
multiagent & $665$ \\
multi scale & $779$ \\
multifractal & $390$ \\
multiscale & $1439$ \\
neural network* & $12747$ \\
\scriptsize{(non linear* OR nonlinear*) NOT equation*} & $10240$ \\
non linear dynamic* & $560$ \\
non linear system* & $391$ \\
nonlinear dynamic* & $2285$ \\
nonlinear system* & $1826$ \\
phase transition & $5503$ \\
plasticity & $6667$ \\
random walk & $758$ \\
robustness & $6498$ \\
scaling & $7008$ \\
social system* & $586$ \\
spin glass* & $643$ \\
\scriptsize{stability AND (lyapunov OR non linear* OR nonlinear*)}  & $1399$ \\
stochastic & $9184$ \\
synchronization & $4645$ \\
turbulence & $4602$ \\
universality & $861$ \\
cell* automat* & $1659$ \\
\hline
\end{tabular}}
\label{TK}
\end{table}

\begin{table*}[!ht]
\caption{
\bf{The 20 references (including books and articles) and journals which are the more used by the articles of the whole database}}
\scriptsize{
\begin{center}
\begin{tabular}{lrclr}
\hline
Reference & Times used & \hspace{5mm} & Journal ($\#$ distinct refs) & Times used \\
\hline
Bak P, 1987,PHYS REV LETT (59) & 2131 && NATURE (29166) & 169309\\
Albert R, 2002, REV MOD PHYS (74) & 2050 && P NATL ACAD SCI USA (42504) & 151140\\
Laemmli UK, 1970, NATURE (227) & 1762 && J BIOL CHEM (59436) & 149042\\
Watts DJ, 1998, NATURE (393) & 1732 && SCIENCE (24880) & 148002\\
Barabasi AL, 1999, SCIENCE (286) & 1693 && CELL (11044) & 99168\\
Bak P, 1988, PHYS REV A (38) & 1555 && PHYS REV LETT (23269) & 94861\\
Sambrook J, 1989, MOL CLONING LAB MANU & 1439 && J AM CHEM SOC (29807) & 82569\\
Newman MEJ, 2003, SIAM REV (45) & 1308 && EMBO J (10926) & 53049\\
Bradford MM, 1976, ANAL BIOCHEM (72) & 1255 && MOL CELL BIOL (12866) & 52694\\
Lowry OH, 1951, J BIOL CHEM (193) & 1106 && J NEUROSCI (12313) & 43152\\
Rumelhart DE, 1986, PARALLEL DISTRIBUTED (1) & 947 && J IMMUNOL (18891) & 41496\\
Strogatz SH, 2001, NATURE (410) & 907 && PHYS REV B (19367) & 41450\\
Kohonen T, 1982, BIOL CYBERN (43) & 901 && J CELL BIOL (10239) & 40560\\
Chomczynski P, 1987, ANAL BIOCHEM (162) & 849 && J CHEM PHYS (17136) & 40074\\
Goldberg DE, 1989, GENETIC ALGORITHMS S & 822 && GENE DEV (4879) & 38903\\
Lorenz EN, 1963, J ATMOS SCI (20) & 726 && BIOCHEMISTRY-US (16035) & 32061\\
Mandelbrot BB, 1982, FRACTAL GEOMETRY NAT & 721 && BRAIN RES (15364) & 30517\\
Kohonen T, 1990, P IEEE (78) & 715  && ANGEW CHEM INT EDIT (7572) & 27718\\
Dorogovtsev SN, 2002, ADV PHYS (51) & 688 && NUCLEIC ACIDS RES (9738) & 27242\\
Albert R, 2000, NATURE (406) & 678 && J EXP MED (8100) & 27220\\
\hline
\end{tabular}
\end{center}
}
\label{mostusedrefs}
\end{table*}

\begin{figure}[!ht]
\begin{center}
\includegraphics[width=0.45\textwidth]{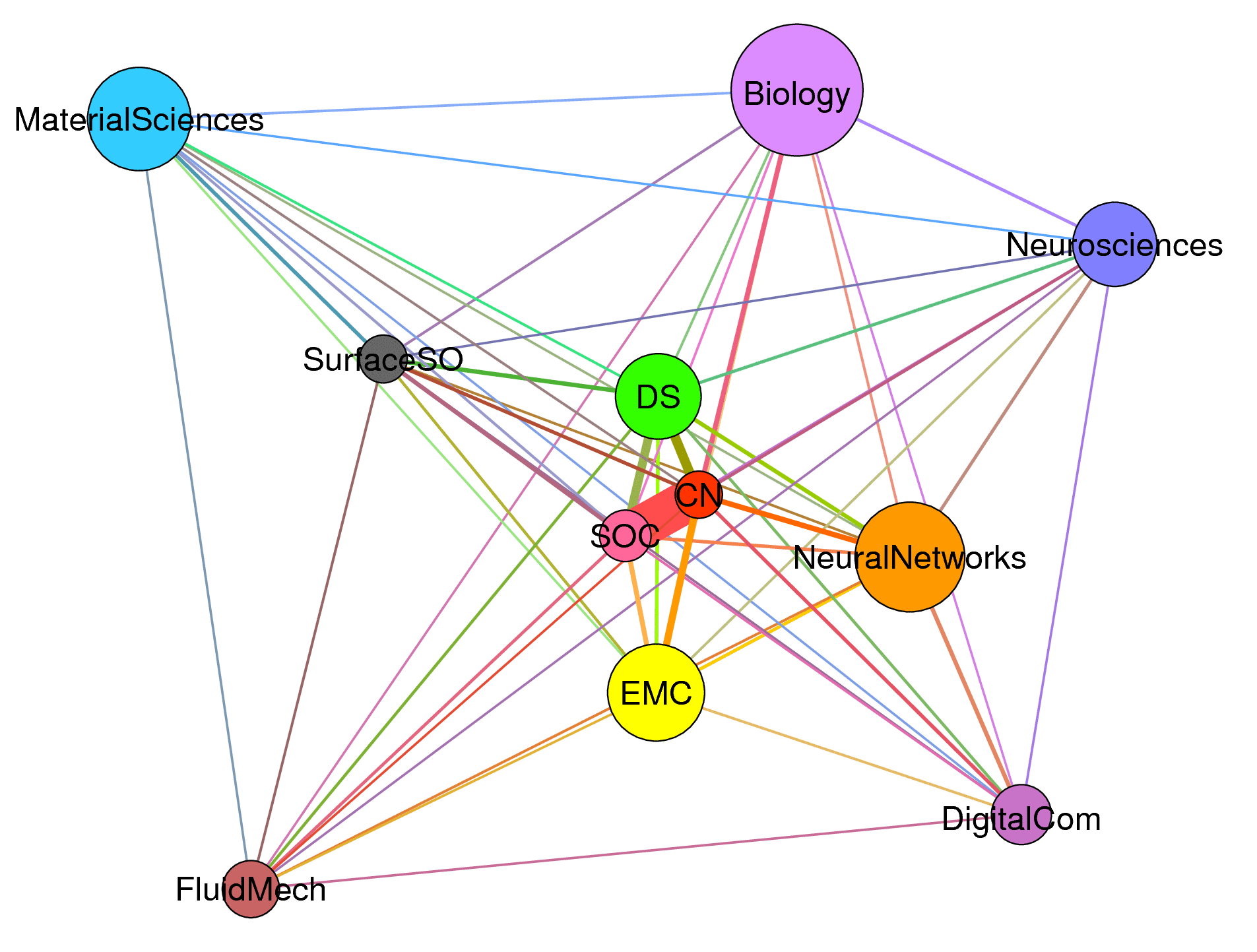}
\end{center}
\caption{
{\bf Community structure obtained with a first run of the modularity maximization \protect\cite{blondel} on the 2000-2008 network ($141\,098$ articles).} The surface of a community $I$ is proportional to its number of articles $N_I$ and the width of the link between two communities $I$ and $J$ is proportional to the mean bibliographic coupling $\langle\omega\rangle_{IJ} = \sum_{i \in I, j \in J} \omega_{ij} / N_I\,N_J$. The layout of the graph is obtained thanks to a spring-based algorithm implemented in the Gephi visualization software \protect\cite{gephi,force}. For the sake of clarity, communities with less than 300 articles are not displayed. The label of a community represents the most frequent and/or significant keyword of its articles. {\it CN} stands for Complex Networks, {\it SOC} for Self Organized Criticality, {\it DS} for Dynamical Systems, {\it DigitCom} for Digital Communication and {\it SurfaceSO} for Self-organization on Surfaces. {\it EMC} is a more composite community where the three most representative keywords are {\it Ecology}, {\it Management} and {\it Computational Models}. See Figure \protect\ref{2000} for details.
}
\label{2000up}
\end{figure}

\begin{figure*}[!ht]
\begin{center}
\includegraphics[width=0.9\textwidth]{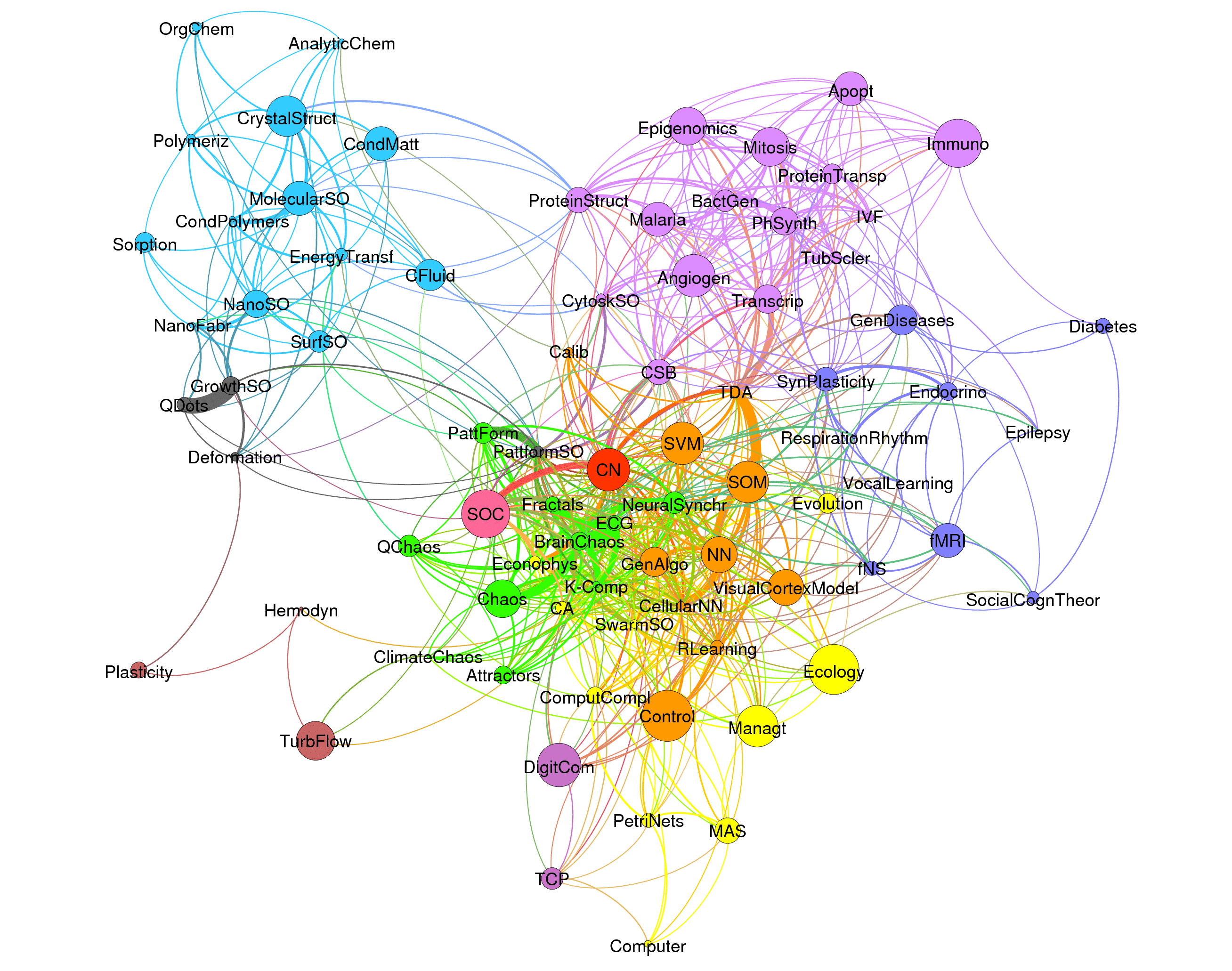}
\end{center}
\caption{
{\bf Community structure obtained with a second run of the modularity maximization on the 2000-2008 network.} This community structure is obtained by optimizing the internal modularity $Q^i$ of each community obtained by the first run of the modularity maximization algorithm on the 2000-2008 network, displayed on Figure \protect\ref{2000up} (See Methods for details on the procedure). The layout of the graph is obtained thanks to a spring-based algorithm implemented in the Gephi visualization software \protect\cite{gephi,force}. The surface of each community is proportional to its number of articles and the width of the link between two communities $I$ and $J$ is proportional to the mean weight $\langle\omega\rangle_{IJ}$. For the sake of clarity, communities with less than 300 articles and links with a mean weight $\langle\omega\rangle_{IJ}$ less than $ 2.10^{-5}$ are not displayed. The color of a community (see online) corresponds to the color of the field (Fig \protect\ref{2000up}) it belongs to. Community labels generally correspond to the most frequent and/or significant keyword. For a detailed presentation of all the subfields, including their authors, most used journals, references and keywords, see the Supplementary Information.
}
\label{2000}
\end{figure*}

\begin{table*}[ht]
\caption{
\bf{2000-2008 subfields' sizes $N$, inner coherences $<\omega>^{-1}$ and modules $q$.}}
\scriptsize{
\begin{center}
\begin{tabular}{l r r r}
\hline
Subfield & $N$ & $<\omega>^{-1}$ & $q$\\
\hline
Analytic Chemistry (AnalyticChem)       & 419 & 336.14 & 0.0002\\
Angiogenesis (Angiogen)                  & 3642 & 1408.78 & 0.005\\
Apoptosis (Apopt)                        & 2632 & 1424.23 & 0.0026\\
Attractors                                & 1161 & 524.42 & 0.0013\\
Bacterial Genomics (BactGen)             & 1517 & 120.19 & 0.0103\\
Brain Chaos                               & 1175 & 70.06 & 0.0105\\
Calibration (Calib)                       & 538 & 371.17 & 0.0004\\
Cellular Automata(CA)                    & 846 & 164.26 & 0.0023\\
Cellular Neural Networks (CellularNN)     & 620 & 86.17 & 0.0024\\
Chaos                                     & 3134 & 531.3 & 0.0099\\
ClimateChaos                              & 352 & 205.31 & 0.0003\\
Complex Fluids (CFluid)                 & 2310 & 994.14 & 0.0029\\
Complex Networks (CN)                     & 3684 & 21.87 & 0.2235\\
Computational Complexity (ComputCompl)  & 1134 & 379.92 & 0.0018\\
Computational Systems Biology (CSB) & 1799 & 323.99 & 0.0053\\
Computer                                 & 526 & 437.08 & 0.0003\\
Condensed Matter (CondMatt)             & 2629 & 631.99 & 0.0059\\
Condensed Matter - Polymers (CondPolymers) & 471 & 131.85 & 0.0009\\
Control                                   & 4772 & 1086.35 & 0.0112\\
Crystal Structure (CrystalStruct)       & 3386 & 350.41 & 0.0175\\
Cytoskeleton Self-Organization(CytoskSO) & 651 & 132.58 & 0.0017\\
Deformation                             & 590 & 432.48 & 0.0004\\
Diabetes                                  & 1015 & 669.24 & 0.0008\\
Digital Communication (DigitCom)          & 3811 & 470.56 & 0.0165\\
Ecology                                  & 4751 & 1846.16 & 0.0066\\
Econophysics (Econophys)                 & 738 & 96.12 & 0.003\\
Electrocardiogram (ECG)                   & 987 & 117.68 & 0.0044\\
Endocrinology (Endocrino)                 & 1223 & 607.63 & 0.0013\\
Energy Transfert (EnergyTransf)         & 801 & 258.27 & 0.0013\\
Epigenomics                              & 3055 & 677.49 & 0.0074\\
Epilepsy                                  & 316 & 273.67 & 0.0001\\
Evolution                                & 1318 & 796.76 & 0.0011\\
Fractals                                  & 1015 & 192.02 & 0.0029\\
Functional MRI (fMRI)                    & 2634 & 897.28 & 0.0041\\
Functional Neurosciences (fNS)           & 935 & 497.44 & 0.0009\\
Genetic Algorithm (GenAlgo)               & 2177 & 197.96 & 0.0128\\
Genetic Diseases (GenDiseases)            & 2273 & 387.34 & 0.0072\\
Growth Self-Organization (GrowthSO)     & 1192 & 113.84 & 0.0067\\
Hemodynamics (Hemodyn)                  & 346 & 344.37 & 0.0001\\
Immunology (Immuno)                      & 4403 & 2234.66 & 0.0047\\
\hline
\end{tabular}
\begin{tabular}{l r r r}
\hline
Subfield & $N$ & $<\omega>^{-1}$ & $q$\\
\hline
In Vitro Fertilization (IVF)             & 591 & 281.01 & 0.0006\\
Kolmogorov Complexity (K-Comp)            & 501 & 121.94 & 0.0011\\
Malaria                                  & 2702 & 747.44 & 0.0052\\
Management (Managt)                      & 3563 & 2159.31 & 0.0031\\
Mitosis                                  & 3171 & 564.98 & 0.0095\\
Molecular Self-Organization (MolecularSO) & 2684 & 409.08 & 0.0094\\
Multi-agent System (MAS)                 & 1787 & 1094.91 & 0.0015\\
Nanofabrication (NanoFabr)              & 457 & 45.28 & 0.0025\\
Nanosciences (Nano)                     & 1995 & 418.01 & 0.0051\\
Neural Networks (NN)                      & 2902 & 221.15 & 0.0201\\
Neural Synchronization (NeuralSynchr)     & 1451 & 453.59 & 0.0025\\
Organic Chemistry (OrgChem)             & 649 & 368.67 & 0.0006\\
Pattern Formation (PattForm)              & 1403 & 205.82 & 0.0051\\
Pattern Formation \& Self-Organization & 691 & 142.82 & 0.0018\\
(PattformSO) && \\
Petri Nets                               & 957 & 275.76 & 0.0018\\
Photosynthesis (PhSynth)                 & 2000 & 224.48 & 0.0096\\
Plasticity                              & 1066 & 915.05 & 0.0006\\
Polimerization (Polymeriz)              & 645 & 98.54 & 0.0022\\
Protein Structure (ProteinStruct)        & 1830 & 237.07 & 0.0076\\
Protein Transport (ProteinTransp)        & 1305 & 308.77 & 0.0029\\
Quantum Chaos (QChaos)                    & 1456 & 636.22 & 0.0018\\
Quantum Dots (QDots)                    & 921 & 130.93 & 0.0035\\
Reinforcement Learning (RLearning)        & 891 & 287.57 & 0.0014\\
Respiration Rhythm                        & 416 & 57.35 & 0.0016\\
Self-Organized Criticality (SOC)         & 4447 & 199.3 & 0.0509\\
Self-Organizing Maps(SOM)                  & 3495 & 168.85 & 0.0376\\
Social Cognition Therory (SocialCognTheor)& 800 & 680.59 & 0.0005\\
Sorption                                & 1354 & 925.37 & 0.001\\
Support Vector Machines(SVM)              & 3660 & 867.91 & 0.0082\\
Surface Self-Organization (SurfSO)      & 1511 & 468.34 & 0.0026\\
Swarm Intelligence (SwarmIntel)          & 608 & 145.94 & 0.0013\\
Synaptic Plasticity (SynPlasticity)       & 1625 & 370.25 & 0.0038\\
Transcriptomics (Transcrip)              & 2043 & 439.37 & 0.0051\\
Transcriptomics Data Analysis (TDA)       & 628 & 43.32 & 0.0049\\
Transmission Control Protocol (TCP)       & 1473 & 718.93 & 0.0016\\
Tuberous Sclerosis (TubScler)            & 766 & 153.33 & 0.002\\
Turbulent Flow (TurbFlow)               & 3172 & 1212.32 & 0.0045\\
Visual Cortex Model                       & 2851 & 845.22 & 0.0051\\
VocalLearning                             & 389 & 162.56 & 0.0005\\
\hline
\end{tabular}
\end{center}
}\normalsize{}
The acronyms and abbreviations in parenthesis correspond to the label of the subfields displayed on Fig \protect\ref{2000}.
The inverse of the average of the weight of the inner links of a subfield $<\omega>^{-1}$ can be taken as an inner coherence measure. Indeed, would the weight of these links be homogeneously distributed between all pairs of articles of a given subfield, then two articles of this subfield chosen at random would share 1 reference over $<\omega>^{-1}$. 
\label{tabledowncomm0009}
\end{table*}

Figure \ref{2000up} shows the largest communities (thereafter also called ``fields'' or ``disciplines'') obtained by modularity maximization of the network of papers published in the years 2000-2008. The layout of all the graphs is obtained thanks to a spring-based algorithm implemented in the Gephi visualization software \cite{gephi,force}. We first note that all important complex systems subfields\footnote{In the following, we use italics to refer to the names of the communities.} are present \cite{cs_sfi}. At the center, we find mostly theoretical domains: {\it self-organized criticality}, {\it dynamical systems}, {\it complex networks}, {\it neural networks}. These fields are connected to more experimental communities lying at the edges ({\it materials science}, {\it biology} or {\it neurosciences}). The links between theoretical and experimental fields suggest that complex systems science models have connections to the ``real'' world, as claimed by their practitioners.

To understand the inner structure of these large communities, we use recursive modularity optimization (see \cite{fortbarth} and ``Methods"). Most fields display a rich inner structure (Figure \ref{2000}) with subcommunities  (thereafter also called ``subfields'' or ``subdisciplines'') organized around specific topics and references. The only exceptions are {\it self-organized criticality} and {\it complex networks}, where all articles cohere around a few references. For a short presentation of all the subfields, see Table \ref{tabledowncomm0009}. For a more detailed presentation of the main subfields, including their authors, most used journals, references and keywords, see the Supplementary Information. We analyze this complex structure at two levels. First, at the global scale, complex systems science appears to be a densely interconnected network. This is somewhat surprising since sharing references between subdisciplines means that they are able to read and understand these references, and moreover, that they find them useful. Would these shared references point to ``universal'' principles? Second, we focus on a more local scale, on the links that specifically connect two different disciplines ({\it ie} two different colors in Figure \ref{2000up}) to understand how they manage to exchange knowledge.

\subsubsection*{Complex systems' science overall coherence}

\begin{figure*}[!ht]
\begin{center}
\includegraphics[width=0.9\textwidth]{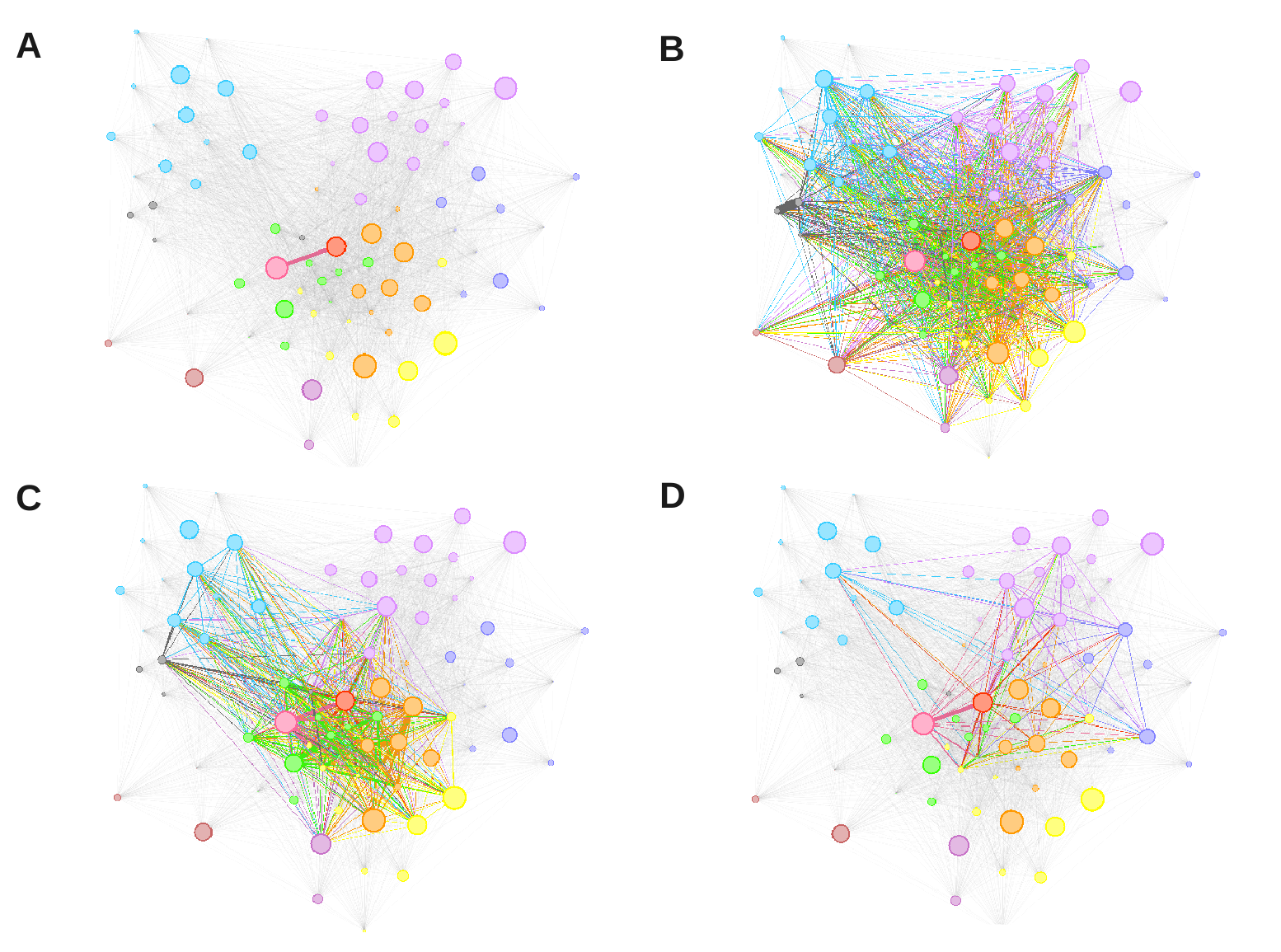}
\end{center}
\caption{
{\bf Local ``networking'' force for four different references on the 2000-2008 network (Fig \protect\ref{2000}). } Links established using the reference are shown in color. The number of citations corresponds to those included in papers of our database published between 2000 and 2008 \textbf{a.} Ref used: Albert \& Barabasi \protect\cite{cn} (2058 citations) \textbf{b.} Ref used: Press WH {\it et al}, Numerical Recipes - all editions \protect\cite{press} (1267 citations) \textbf{c.}  Ref used: Nicolis \protect\cite{nicolis} (342 citations) \textbf{d.} Ref used: Barabasi and Oltvai \protect\cite{nrg} (244 citations).
}
\label{net}
\end{figure*}

Let us start with the field's overall coherence. We have looked for the references cited by many subfields. These form the ``glue'' that links many subdisciplines and connects the network. More precisely, we define the networking force of a reference  $\mathcal{N} (r)$ as the sum, over all pairs of subfields, of the proportion of their links explained by that reference (see Methods). Table \ref{glue} shows that the references that glue the network are more methodological than theoretical: the most networking reference is ``Numerical Recipes'' \cite{press}, a series of books that gathers many routines for various numerical calculations and their implementation in computers. Most of the other linking references are mathematical handbooks or data analysis tools. If one looks for universality in the complex systems field, the computer -- as a tool -- seems to be a serious candidate. Among the leading contributors to the glue, we also find several references on self-organization (SO). Self-organization is not a predictive theory, but an approach that focuses on the spontaneous emergence of large-scale structures out of local interactions between the system's subunits \cite{mehdi}. Several subdisciplines in Figure \ref{2000} can be related to this approach, as they use a keyword akin to ``self-organization'' (SO) in more than 10\% of their articles (for a more complete list of the communities using this keyword, see the Supplementary Information). Among these we find {\it swarm SO}, {\it molecular SO} linking chemistry to biology, {\it growth SO} and {\it pattern formation SO} linking surface science to dynamical systems. This suggests that the field of complex systems focuses on the cases in which the link from microscopic to macroscopic can be analyzed through self-organization, which gave rise to several fruitful scientific programs, as we discuss below (section Discussion).

\subsubsection*{Interdisciplinary trading zones}

At a more local scale, let us now look at the links that specifically connect two distinct disciplines. How are those connections established? It is widely accepted that scientific disciplines cannot easily communicate or be linked (in our case, share references) simply because it is difficult for a physicist to understand a biology paper and vice-versa. In addition, different disciplines have different definitions of what counts as a result or as an interesting research topic. For example, physical sciences look for universal laws, while social \cite{borgatti} and biological \cite{efk_power} sciences emphasize the variations in structure across different groups or contexts and use these differences to explain differences in outcomes. Physicians are interested in practical medical advances while physicists want to know whether physiological rhythms are chaotic or not \cite{glass}.

\begin{figure}[!ht]
\begin{center}
\includegraphics[width=0.4\textwidth]{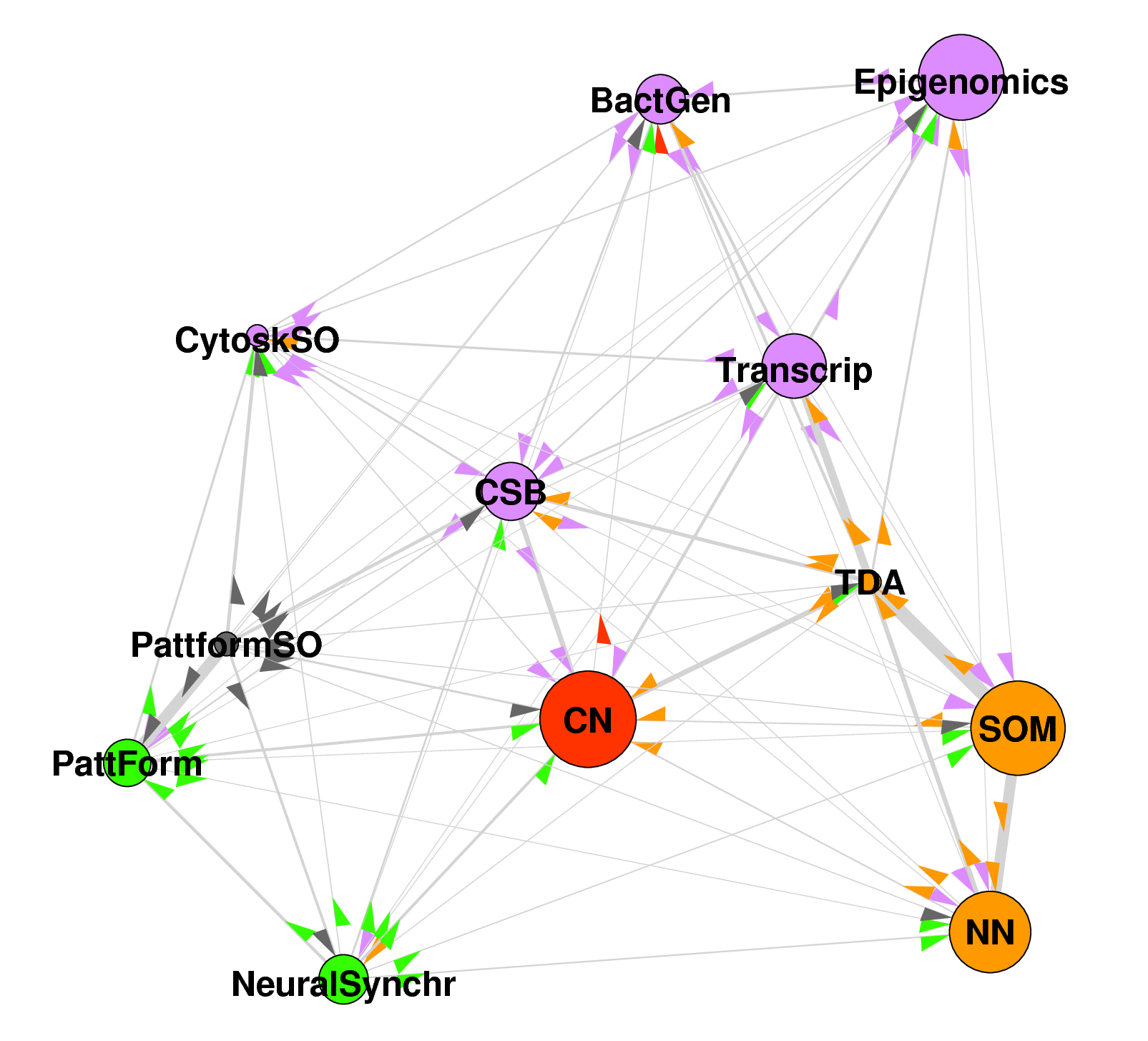}
\end{center}
\caption{
{\bf Directed network.}  On this subset of the graph presented in Figure \ref{2000}, the arrows are directed to the subfield that uses the other subfield's references to establish the link. More precisely, the common references shared by two linked subfields are more similar to the internal references of the subfield from which the arrow originates than to the internal references of the subfield to which the arrow points (see ``Methods'' for more details). The figure shows that {\it transcriptional data analysis} (TDA) feeds from {\it self-organizing maps} (SOM) and {\it neural networks} (NN) methodological references, while biology subcommunities (mainly Transcriptomics) use {\it transcriptional data analysis} references. The orientation of the links is quite different for {\it computational systems biology} (CSB) and {\it complex networks} (CN), because these subfields tend to pump their neighbors' references, while the other subfields do not find much use in {\it computational systems biology} and {\it complex networks} references.
}
\label{TZ}
\end{figure}

Where do the links come from then? In an illuminating analogy, Peter Galison \cite{galison} compares the difficulty of connecting scientific disciplines to the difficulty of communicating between different languages. History of language has shown that when two cultures are strongly motivated to communicate - generally for commercial reasons - they develop simplified languages that allow for simple forms of interaction. At first, a ``foreigner talk'' develops, which becomes a ``pidgin'' when social uses consolidate this language. In rare cases, the ``trading zone'' stabilizes and the expanded pidgin becomes a creole, initiating the development of an original, autonomous culture. Analogously, biologists may create a simplified and partial version of their discipline for interested physicists, which may develop to a full-blown new discipline such as biophysics. Specifically, Galison has studied \cite{galison} how Monte Carlo simulations developed in the postwar period as a trading language between theorists, experimentalists, instrument makers, chemists and mechanical engineers. Our interest in the concept of a trading zone is to allow us to explore the dynamics of the interdisciplinary interaction instead of ending analysis by reference to a ``symbiosis'' or ``collaboration''. 

\begin{table}[!ht]
\caption{
\bf{The 20 most networking references in the 2000-2008 decade}}
\begin{center}
\begin{tabular}{llr}
\hline
Reference & Topic & $\mathcal{N} (r)$ (\%)\\
\hline
Press et al. (1992)*  & Numerical recipes (book)  & 1.250\\
Shannon (1948)*             & Information theory        & 0.607\\
Metropolis et al. (1953)    & Monte Carlo integration   & 0.509\\
Nicolis et al. (1977)* & Self organization (book)  & 0.420\\
Kauffman (1993)*            & Self organization (book)  & 0.309\\
Hebb (1949)                 & Neuropsychology and behavior       & 0.297\\
&  (book) &\\
Alberts et al. (1994) & Molecular and cellular biology      & 0.288\\ 
&  (book) &\\
Abramowitz et al. (1968)* & Handbook of mathematical   & 0.269\\
& functions & \\
Feller (1958)*              & Introduction to probability  & 0.268\\
& theory (book) &\\
Watson \& Crick (1953)       & Structure of DNA          & 0.250\\
Lakowicz (1999)             & Fluorescence spectroscopy & 0.249\\
Turing (1952)               & Morphogenesis             & 0.237\\
Witten et al. (1981)      & Diffusion-limited aggregation             & 0.234\\
Cohen (1988)                & Statistics and behavioral  & 0.223\\
& sciences (book) & \\
Hopfield (1982)             & Neural networks           & 0.217\\
Stanley (1971)              & Phase transition (book)   & 0.202\\
Whitesides et al. (2002)  & Self-assembly         & 0.188\\
Marquardt (1963)            & Applied mathematics  & 0.174\\
Chomczynski (1987)     & RNA isolation             & 0.167\\
Venter et al. (2001)        & Human genome sequence     & 0.160 \\
\hline
\end{tabular}
\end{center}
\begin{flushleft}
The references followed by a star correspond to books or papers which appeared in the database under several forms - essentially different publication years for the books - for which the networking power $\mathcal{N} (r)$ have been summed. The complete references of these papers are given in Supplementary Information.
\end{flushleft}
\label{glue}
\end{table}

Table \ref{TZ0009} gives a list of the main ``trading zones'' which connect theoretical and experimental fields in Figure \ref{2000up} and capture a significant fraction of the links between these fields. The clearest example is {\it transcriptomics data analysis}, a subfield of {\it neural networks} which connects biologists interested in the interpretation of data retrieved from DNA chips and computer scientists interested in data analysis via methods from the {\it neural networks} field. The {\it transcriptomics data analysis} subfield represents 2.3\% of {\it neural networks} papers but accounts for 46.3\% of the connections between {\it neural networks} and {\it biology} and 16.5\% of the links between {\it neural networks} and {\it complex networks}. Other trading zones are {\it computational systems biology}, linking {\it biology} to many theoretical fields, among which {\it dynamical systems}, {\it self-organized criticality} and {\it complex networks}, {\it neural synchronization} linking {\it dynamical systems} and {\it neurosciences}, {\it cytoskeleton self-organization} linking {\it biology} to {\it dynamical systems} and {\it self-organized criticality} and {\it calibration} linking {\it neural networks} and {\it material sciences}. Note that a single trading zone can be used by a fields to exchange with several other fields, as long as these other fields share the same ``language''. For example, {\it computational systems biology}, links {\it biology} to {\it dynamical systems}, {\it self-organized criticality} and {\it complex networks}, three subfields which share the physicists' toolkit. Since our map cannot cover all scientific fields, we may not recognize some subfields as trading zones, such as {\it electrocardiogram} which is likely to connect {\it dynamical systems} to medecine, or even miss a trading zone between geosciences and {\it self-organized criticality}.

\begin{table}[!ht]
\caption{
\bf{Strongest trading zones.}}
\begin{center}
\scriptsize{
\begin{tabular}{llrr}
\hline
Subfield & Fields & $T$ ($\%$) & $T/T^{exp}$\\
\hline
TDA & Biology/Neural Networks & 46.355 & 20.51\\
CSB & Biology/Dynamical Systems & 49.704 & 8.87\\
CSB & Biology/SOC & 49.255 & 8.79\\
ProteinStruct & Biology/Material Sciences & 47.442 & 8.32\\
CSB & Biology/CN & 42.931 & 7.66\\
TDA & CN/Neural Networks & 16.543 & 7.32\\
Hemodyn & Neurosciences/FluidMech & 54.552 & 7.22\\
NeuralSynchr & Neurosciences/Dynamical Systems & 59.788 & 5.20\\
Hemodyn & Biology/FluidMech  & 39.113 & 5.18\\
CytoskSO & Biology/SOC & 9.561 & 4.71\\
CytoskSO & Biology/Dynamical Systems & 9.522 & 4.69\\
Calib & Material Sciences/Neural Networks & 8.717 & 4.51\\
CellularNN & CN/Neural Networks & 9.605 & 4.30\\
Transcrip & Biology/CN & 25.806 & 4.05\\
\hline
\end{tabular}}
\end{center}
\begin{flushleft}
The trading force $T$ of a subcommunity measures the fraction of the links between two fields (in Fig. \protect\ref{2000up}) which goes through this subcommunity. More precisely, the trading force of $I$, a subcommunity of $\bar{I}$, towards any community $\bar{J}$ is the total weight of the article-article links between the subcommunity $I$ and community $\bar{J}$, normalized by the total weight of the the article-article links between $\bar{I}$ and $\bar{J}$: $T_{\bar{I}\bar{J}}(I) = \sum_{i \in I,j\in \bar{J}} \omega_{ij} / \sum_{i \in \bar{I},j\in \bar{J}} \omega_{ij}$. The expected force $T^{exp}$ is the value of the trading force one would expect if all the links between $\bar{I}$ and outside communities were equally shared among all sub-communities of $\bar{I}$, which is simply the fraction $N_I/N_{\bar{I}}$ of articles of $I$ in $\bar{I}$. The acronyms of the subfiels used here correspond to those explained in Table \protect\ref{tabledowncomm0009}.
\end{flushleft}
\label{TZ0009}
\end{table}

By analyzing carefully the references used by trading zones and also the references that make the links between the trading zones and their neighbors, we can distinguish two types of trading zones, applicative and speculative. Let us start with {\it transcriptomics data analysis}, which is a clear example of ``applicative" trading zone. The development of  new measurement techniques in cellular biology (mainly DNA microarrays) produced huge amounts of data together with the need of new tools to analyze them. Since this new technique promised a better understanding of cell dynamics, a new scientific subdiscipline, able to understand data analysis and its biological interest was built around transcriptomic tools. The two references most used by this subfield stress the applicative side: the purpose of the first paper is ``{\em to describe a system of cluster analysis for genome-wide expression data from DNA microarray hybridization [\ldots] in a form intuitive for biologists}'' \cite{eisen} while the second ``{\em describes the application of self-organizing maps for recognizing and classifying features in complex, multidimensional [transcriptomic] data}'' \cite{tamayo}. The {\it transcriptomics data analysis} papers are clustered together because they share references presenting this kind of applications. The applicative character of {\it transcriptomics data analysis} can also be seen in the origin of the references that link them to neighbor subfields (Figure \ref{TZ}). The common references between {\it transcriptomics data analysis} and {\it biology} (mainly {\it transcriptomics}) are similar to references used by {\it transcriptomics data analysis} papers themselves. This means that the link arises from biologists citing results obtained by {\it transcriptomics data analysis} scientists or techniques they use. On the other hand, the common references between {\it transcriptomics data analysis} and {\it self-organizing maps} (a subfield of {\it neural networks}) are similar to references used by {\it self-organizing maps} papers. Therefore, the link arises from {\it transcriptomics data analysis} scientists citing classification techniques created by {\it self-organizing maps} scientists, while these scientists do not often use {\it transcriptomics data analysis} references. Therefore, {\it transcriptomics data analysis} allows {\it self-organizing maps} techniques to be understood and used to interpret biological data, with a relevance certified by biologists' citations. 
The case of another trading zone, {\it computational systems biology}, is different. Its most used references point to computational methods - mainly Gillespie's algorithm \cite{gillespie} or to experimental papers in which there is no explicit modeling but that show complex cellular dynamics, thus justifying indirectly the need for modeling. The link between experiments and modeling is still speculative, as summarized by one of the most used references in this subfield \cite{tyson}: ``{\em we hope that this review will [...] promote closer collaboration between experimental and computational biologists.}'' Moreover, the common references between {\it computational systems biology} and {\it biology} are from {\it biology}, as if {\it computational systems biology} scientists were eager to quote potentially interesting biological applications for their modeling approach, while many biologists were still unaware of these models. In short, compared to {\it transcriptomics data analysis}, {\it computational systems biology} seems a more speculative trading zone, at the frontier of biology and modeling, but presently lacking a specific object or concept to define an operational trading zone.

\section*{Discussion}
\label{discussion}

Our empirical study of the ``complex systems'' field shows that its overall coherence does not arise from a universal theory but from computational techniques and successful adaptations of the idea of self-organization. The computer is important for advancing the understanding of complex systems because it allows scientists to play with simple but nonlinear models and to handle large sets of data obtained from complex systems. At a more local level --- specifically the interdisciplinary level --- trading zones allow for coordination between vastly different scientific cultures, who differ on their conception of an interesting topic, but who can work together around specific tools (a DNA microchip) or concepts (a network).

We now discuss how our study sheds light on the overall philosophy of the complex systems field. First, we examine the various claims to universality. A ``general systems theory'' would possess a collection of theoretical books or papers revealing the ``universal'' explanation; this would be evidenced in Figure~\ref{2000} by a central group to which other groups would connect. Instead, our analysis shows a variety of modeling disciplines in a central position.

We argue that claims to universality are part of a rhetoric that legitimates the study of abstract and simple models \cite{kids}. Certainly, a few theoretical papers, such as Bak's \cite{bak} (in SOC) or Albert and Barabasi's \cite{cn} (in CN) point to ``universal'' mechanisms and are heavily cited. However, more than 90\% of their citations arise from modelers themselves \cite{woscite}, suggesting that they may be universal... for theorists. Our data support the local character of these ``universal'' laws. First, Albert and Barabasi's \cite{cn} paper is the most cited in the 2000-2008 decade but only links {\it complex networks} and {\it self-organized criticality} subfields (Figure \ref{net}a). The contrast with the global networking achieved by methodological references such as Metropolis' algorithm or Numerical Recipes (Figure \ref{net}b) or self-organization references (Figure \ref{net}c) is clear. Second, the references that {\it complex networks} (Figure \ref{TZ}) and {\it self-organized criticality} communities share with experimental fields are similar to those of the experimental fields. This seizure of experimental references suggests that the links between modeling practices and their potential applications are mostly rhetorical: {\it complex networks} and {\it self-organized criticality} papers often quote experimental work as legitimating their models, while experimentalists rarely refer to them. To try to become universal, theoretical approaches have to be ``translated" into other disciplines. An example of this strategy is shown in Figure \ref{net}d which shows the links established between network science and biology thanks to Barabasi and Oltvai's introduction of networks for biologists \cite{nrg}. Regardless, many physicists are likely to continue looking for common patterns across systems to justify their neglect of the ``details'' of the system under study, the precise components and interactions \cite{efk_power,stanley}. Universality is then another name for simplicity, a strong motivation for many physicists as expressed by the Santa Fe institute who aims at uncovering ``the mechanisms that underlie the deep simplicity present in our complex world.'' \cite{sfi_simple}. It is true that simple analytical models such as SOC or chaotic systems may lead to complicated behaviors and patterns. But this does not prove the reverse proposition, i.e. that all complex patterns can be explained by simple mechanisms. The ``simplicity'' approach turned out to be a successful strategy in the study of phase transitions, which can be studied through the very simple Ising model \cite{efk,castellano,solebak}, but arguments for the usefulness of such an approach for biological or social systems  are unconvincing \cite{kids}.

It could be argued that links between these theories and experimental fields take time to establish and will be seen in the future. An interesting insight of the possible evolution of universality claims is given by the history of self-organization, which was considered by many as a universal key to Nature in the 1980's \cite{efk}. This idea was fecund in that it gave birth to several active subdisciplines ({\it cytoskeleton SO}, {\it growth SO}...) (Figure \ref{2000}). However, it should be noticed that these heirs of self-organization are nowadays almost unrelated. The different {\it self-organization} subfields are more linked to their own discipline ({\it biology}, {\it materials science} \ldots) than between them. This is shown by the plain fact that community detection puts these SO subfields into different disciplines (different colors in Figure \ref{2000}) instead of creating a single, unified, {\it self-organization} field. The reason is that these subfields use widely different references, as illustrated by the fact that there is no common reference among the 10 most used references for all the different {\it self-organization} subcommunities. Self-organization is therefore not a universal explanation but rather a kind of banner, which needs to be associated to references to specific elements (including techniques, microscopic entities and their interactions) to be fruitful.

To understand the essential role of the computer, it is important to distinguish between complex and complicated. A complex phenomenon has to be understood synthetically, as a whole, while a complicated phenomenon can be explained analytically \cite{venturini}. In other words, a system is complex when the ``parts'' that are relevant to link the micro to the macro cannot be properly defined, as in social systems where humans cannot be defined without taking into account society (e.g. language, an essential part of the individual which is acquired through society). Similarly, synthetic biology aims at using functional components of living systems as building blocks to create artificial devices \cite{benner}. But many difficulties arise from the intertwining of the elements in a living being \cite{serrano}. Biological parts are ill defined and their function cannot be isolated from the context that they themselves create. Despite its claims to complexity and holism, the ``complex systems'' field proposes a standard mechanistic vision of nature and society. As for most natural sciences, its aim is to transform complex systems into complicated systems that can be handled and eventually engineered by models and computer force.

This is confirmed by historical studies showing \cite{schweber,efk} that the complex systems field is heir of the postwar sciences born around the computer: operational research, game theory and cybernetics. These fields started when physicists, mathematicians and engineers started collaborating to maximize the efficiency of WW II military operations \cite{mangle,schweber,bowker}. These sciences extended the mechanistic, engineering vision of the physical world to the biological and social worlds. This view is still present in many today's prominent CS scientists: ``our knowledge of [social] mechanisms [..] is essential for self-optimization of the society as a whole'' \cite{palla}; ``We spend billions of dollars trying to understand the origins of the universe, while we still don't understand the conditions for a stable society, a functioning economy, or peace'' \cite{helbing} or ``[Systems biology] leads to a future where biology and medicine are transformed into precision engineering'' \cite{kaneko}.

\bigskip
In summary, we have obtained a {\it global} point of view on the structure of the "complex systems" field. This has allowed us to test empirically the idea of universality, showing that it remains a dream, albeit one which has lead to interesting but more modest realities. At the global scale, the whole domain is linked by the focus on self-organization and the use of computer-based methods for solving non-linear models. At a more local scale, the links between different disciplines are achieved through the development of ``trading zones'' \cite{galison}. These allow for coordination between vastly different scientific cultures, for example theoretical and experimental disciplines, which are only marginally connected. These disciplines may differ on the very conception of what is an interesting topic, but can work together around specific tools (a DNA microchip) or concepts (a network). Today, these interdisciplinary collaborations are a key to essential scientific challenges such as the analysis of the massive amount of data recently made available on biological and social systems \cite{social_science,microsoft} and the understanding of the complex intertwining of different levels of organization that is characteristic of these systems.

\section*{Methods}

\subsubsection*{Extraction of the data}
Our data have been extracted from the ISI Web of Knowledge database \cite{wos} in December 2008. The science of complex systems is particularly challenging as an epistemic object since there exists no consensual definition of the domain, nor any list of disciplines or journals that would gather all the relevant papers. Therefore, we selected all the articles of the database whose title, abstract (for articles published after 1990) or keywords contained at least one of a chosen list of topic keywords (Table \ref{TK}). These keywords were derived from discussions with experts of the field, mainly scientists working at the complex systems institute in Lyon (IXXI). We have retrieved $215\,305$ articles ($141\,098$ between 2000 and 2008) containing $4\,050\,318$ distinct references. Each record contains: authors, journal, year of publication, title, keywords (given by the authors and/or ISI Web of Science) and the list of references of the article. Any choice of keywords being potentially biased and partial, our strategy was to risk choosing too many of them - thus bypassing the lack of precise definition of the ``complex systems'' field and retrieving all its important subfields - and to trust the subsequent analysis to eliminate irrelevant articles.

In fact, as shown in table \ref{TK}, around $40\%$ of the articles of the database comes solely from the combination of keywords {\it``complex*'' and ``control''}. While most of those articles were close to biology and not directly related to the field of complex systems, we chose to keep them in order to test the robustness of our analysis. As shown below, our strategy was successful, since most of these ``irrelevant'' articles are grouped into a few communities (such as {\it Apoptosis} or {\it Immunology}) that lie at the network's edges and do not bias the results.

\subsubsection*{Links between articles}

Weight of links between articles are calculated through their common references (bibliographic coupling \cite{kessler}). We define a similarity between two articles $i$ and $j$ as the cosine distance: 

\begin{equation}
\omega_{ij} = \frac{|\mathcal{R}_i\cap \mathcal{R}_j|}{\sqrt{|\mathcal{R}_i|\,|\mathcal{R}_j}|}
\label{cosine}
\end{equation}

where $\mathcal{R}_i$ is the set of references of article $i$. By definition, $\omega_{ij} \in [0,1]$, is equal to zero when $i$ and $j$ do not share any reference and is equal to $1$ when their sets of references are identical. For this study, bibliographic coupling offers two advantages over the more usual co-citation link: it offers a faithful representation of the fields, giving equal weight to all published papers (whether cited or not) and it can be applied to recent papers (which have not yet been cited). Moreover, the links are established on the basis of the author's own decisions (to include or not a given reference) rather than retrospectively from other scientists' citations. Thus, bibliographic coupling can be used to analyze the community of research as it builds itself rather than as it is perceived by later scientists that cite its publications.

\subsubsection*{Community detection and characterization}
In order to structure this network into groups of cohesive articles, we partition the set of papers by maximizing the modularity function. Given a partition of the nodes of the network, the modularity is the number of edges inside clusters (as opposed to crossing between clusters), minus the expected number of such edges if the network was randomly conditioned on the degree of each node. Community structures often maximize the modularity measure. We compute our partition using the algorithm presented in \cite{blondel}, which is designed to efficiently maximize the modularity function in large networks. More precisely, we used the weighted modularity $Q$ \cite{newman,fortunato}, which is defined as $Q = \sum_I q_I$, where the {\em module} $q_I$ of a community $I$ is given by 

\begin{equation}
q_I=\frac{\Omega_{II}}{\Omega} - \left( \frac{\sum_{J\neq I}\Omega_{IJ} + 2 \Omega_{II}}{2\Omega}\right)^2  
\label{modularity}
\end{equation}

where
\begin{eqnarray*}
\Omega_{II} &=& \frac{1}{2}\sum_{i \in I,\,j \in I} \omega_{ij} \\ 
&& \mbox{is the total weight of the links inside community $I$,}\\ 
\Omega_{IJ} &=& \sum_{i \in I,\,j \in J}\omega_{ij}\\ 
&& \mbox{is the total weight of the links between}\\ 
&& \mbox{ communities $I$ and $J \neq I$}\\ 
\Omega &=& \frac{1}{2}\sum_{i,j} \omega_{ij}=\sum_{(i,j)} \omega_{ij}\\ 
&& \mbox{is the total weight of the links of the graph.}
\end{eqnarray*}

Each module $q_I$ compares the relative weight of edges $\frac{\Omega_{II}}{\Omega}$ inside a community $I$ with the expected weight of edges $\left( \frac{\sum_{J\neq I}\Omega_{IJ} + 2 \Omega_{II}}{2\Omega}\right)^2$ that one would find in community $I$ if the network were a random network with the same number of nodes and where each node keeps its degree, but edges are otherwise randomly attached. See Ref \cite{fortbarth} for a more explicit interpretation of the modularity, its properties and limits.

Applying the Louvain algorithm yields a first partition of the network into communities (also referred to as ``fields'' or ``disciplines'', see Figure \ref{2000up}). To obtain the substructure of these communities, we apply the Louvain algorithm a second time on each of them. We find that most of these communities display a clear substructure with high values of internal modularity $Q^i$ (typically between $0.4$ and $0.8$). Only two of them ({\it self-organized criticality} and {\it complex networks}) are strongly bound around a few references and present much lower values of $Q^i$ (typically less than or around $0.2$). Consequently they were not split into subfields which would not have much scientific relevance.

This recursive modularity optimization \cite{fortbarth} leads us to a ``subfield'' graph (Figure \ref{2000}). We have checked that all the obtained sub-communities satisfy the criterion ($q_I \geq  0$) proposed by Fortunato and Barth\'el\'emy \cite{fortbarth} to check their relevance (see Table \ref{tabledowncomm0009}). 

\subsubsection*{Links between communities and their orientation}

The link between two communities $I$ and $J$ can be quantified by the average distance between an article $i \in I$ and an article $j \in J$:
\begin{equation}
<w>_{IJ}^{-1}\, = \, <w_{ij}>_{i \in I, j \in J}^{-1} \, = \, \left(\Omega_{IJ} / N_I\,N_J\right)^{-1} 
\end{equation}

A link between a community $I$ and a community $J$ exists if at least one reference is shared between an article of $I$ and an article of $J$. To analyze the scientific content conveyed by the link, it is important to know if the shared references are more similar to the references used by community $I$ or to the references used by community $J$. To take into account this similarity, we define the {\em orientation} of a community-community link in the following way.

Let $n_{r,I}$ be the number of papers of community $I$ using reference $r$. Then,
\begin{itemize}
\item the number of article-article links inside community $I$ which use reference $r$ is $L_{r,II} = n_{r,I}\,(n_{r,I}-1)/2$
\item the number of article-article links between communities $I$ and $J$ which use reference $r$ is $L_{r,IJ}=n_{r,I}n_{r,J}$
\end{itemize}

We compare the set of references shared by the two communities $I$ and $J$ to the references used by $I$ and $J$ by computing the cosine similarity measures:
\begin{eqnarray}
\cos_{II,IJ} &=& \frac{\sum_r L_{r,II}L_{r,IJ}}{\sqrt{\sum_r L_{r,II}^2 \sum_r L_{r,IJ}^2}} \\
&& \mbox{comparing the shared refs to those of $I$}\nonumber\\
\cos_{JJ,IJ} &=& \frac{\sum_r L_{r,JJ}L_{r,IJ}}{\sqrt{\sum_r L_{r,JJ}^2 \sum_r L_{r,IJ}^2}} \\
&& \mbox{comparing the shared refs to those of $J$} \nonumber
\end{eqnarray}

For example, if $\cos_{II,IJ} < \cos_{JJ,IJ}$, the shared references are more similar to the references binding community $J$ than to the references binding community $I$. We then direct the link from community $J$ to community $I$, as community $I$ ``pumps" community $J$ references to establish the link. See Figure \ref{TZ} for examples of link orientation. 

\subsubsection*{Visualizing linked communities}

To obtain Figures \ref{2000up} and \ref{2000}, we use Gephi \cite{gephi}. The layout of the graph is obtained thanks to a spring-based algorithm implemented in it \cite{force}. ForceAtlas is a force directed layout: it simulates a physical system. Nodes repulse each other (like magnets) while edges attract the nodes they connect (like springs). These forces create a movement that converges to a balanced state, which helps in the interpretation of the data.

\subsubsection*{Networking power of references}

To understand which references link the different subdisciplines to form a connected network, we define the ``glue'' as the set of references shared between subfields. To give equal weight to all these links, we normalize each link to 1, leading to the normalized networking strength $\mathcal{N} (r)$ of reference $r$ as:

\begin{equation}
\mathcal{N} (r) = \frac{1}{Z}\sum_{I \neq J} f_{IJ}(r)
\end{equation}

\noindent where $f_{IJ}(r)$ is the fraction of links between an article of community $I$ and an article of community $J$ in which reference $r$ is used and where $Z$ is a normalization constant such that $\sum_r \mathcal{N} (r) = 1$. The normalization ensures that $\mathcal{N} (r)$ represents the proportion of all the links of the complex systems field that can be assigned to reference $r$.

\section*{Acknowledgments}
We acknowledge interesting discussions with Andre\"i Mogoutov, L\'eo Granger, Taras Kowaliw and Eric Bertin.


\end{document}